\documentclass[%
 reprint,
 amsmath,amssymb,
]{revtex4-2}

\usepackage{graphicx}
\usepackage{dcolumn}
\usepackage{bm}
\usepackage{braket}
\usepackage{lipsum}

\begin{document}


\title{Compensation of the multipolar polarizability shift in optical lattice clocks}

\author{Artem Golovizin}
\email{artem.golovizin@gmail.com}
\affiliation{P.N.\,Lebedev Physical Institute, Leninsky prospekt 53, 119991 Moscow, Russia}
\affiliation{Russian Quantum Center, Bolshoy Bulvar 30,\,bld.\,1, Skolkovo IC, 121205 Moscow, Russia}

\date{\today}

\begin{abstract}

In neutral atom optical clocks, the higher-order atomic polarizability terms lead to the clock transition frequency shift which is motion-state dependent and nonlinear with the optical lattice depth. 
We propose to use an auxiliary optical lattice to compensate the influence of the E2-M1 differential polarizability or tune the associated coefficient to a favorable value. 
We show that by applying this method to Sr and Hg optical lattice clocks, the low or even sub-$10^{-19}$ clock transition frequency uncertainty from the optical lattice becomes feasible.
Finally, the proposed scheme is simple for experimental realization and can be implemented and tested in the existing setups. 
\end{abstract}

\maketitle

Optical clocks became extremely precise instruments surpassing microwave frequency standards by two orders of magnitude in frequency accuracy and stability \cite{bothwell2019jila, brewer2019al+, takamoto2022perspective, mcgrew2018atomic}. 
As a result, redefinition of the SI unit of time, the second, based on an optical transition is scheduled for 2030  \cite{dimarcq2023roadmap}. 
Furthermore, the capability to control and compare optical frequencies at $10^{-18}$ level or better allows stringent tests of fundamental theories like general relativity \cite{takamoto2020test, bothwell2022resolving}, Lorentz violation \cite{sanner2019optical}, fundamental constants variation \cite{lange2021improved} and others \cite{derevianko2022fundamental, safronova2018search}.

Optical lattice clocks demonstrate remarkable measurement precision \cite{oelker2019demonstration, bothwell2022resolving} owing to interrogation of a large number of atoms, typically a few thousands. 
At the same time, optical lattice is one of the most significant sources of the clock transition frequency shift and uncertainty.
After pioneering proposal of the concept of magic-wavelength optical lattice \cite{katori1999optimal}, more and more accurate consideration of the optical lattice effects became necessary to provide control of the lattice-induced clock transition frequency shift at the $10^{-17}$ level and better \cite{katori2009magic, brown2017hyperpolarizability, porsev2018multipolar, ushijima2018operational, beloy2020modeling, wu2023contribution, porsev2023contribution, kim2023evaluation}.
This includes accounting for the motion of atoms in the trap and higher order terms of the dynamic atomic polarizabilities $\alpha^\textrm{M1}$ and $\alpha^\textrm{E2}$ associated with magnetic dipole (M1) and electric quadrupole (E2) transitions, respectively, and hyperpolarizability $\beta$.
Even though they are much smaller than the electric dipole polarizability $\alpha^\textrm{E1}$ (typically $\alpha^\textrm{qm}/\alpha^\textrm{E1} \sim 10^{-6}$, where $\alpha^\textrm{qm}=\alpha^\textrm{M1}+\alpha^\textrm{E2}$), their contribution can not be neglected due to different spatial dependence in the optical lattice.

For a one-dimensional optical lattice, the trapping potential for a state $\xi$ ($g$ for the ground or $e$ for the clock states) can be written as \cite{ushijima2018operational}
\begin{equation}
    U_\xi = -\alpha^\textrm{E1}_\xi I \cos^2(kz) - \alpha^\textrm{qm}_\xi I \sin^2(kz) - \beta_\xi I^2 \cos^4(kz) \label{eq:u_s}
\end{equation} 
since the E1-polarizability and hyperpolarizability terms are proportional to $E^2$ and $E^4$, respectively, but M1 and E2 polarizability terms are proportional to $B^2$ and $(\nabla E)^2$, which are both shifted by $\lambda_\textrm{L}/4$ in the standing wave.
Here $k= 2\pi/\lambda_\textrm{L}$ is the wavevector, $\lambda_\textrm{L}$ is the optical lattice wavelength, $I$ is the peak intensity, $E$ and $B$ are the electric and magnetic field amplitudes.
We consider the case when the lattice and probe beam are oriented along the $z$-axis.

For spectroscopy in the Lamb-Dicke regime, the lattice-induced clock transition frequency shift $\Delta\nu_\textrm{LS}$ can be expressed as \cite{ushijima2018operational}:

\begin{equation}
\begin{split}
    h\Delta\nu_\textrm{LS}(u,\delta_\textrm{L},n_\textrm{z}) &= \left(\frac{\partial\tilde{\alpha}^\textrm{E1}}{\partial\nu}\delta_\textrm{L} - \tilde{\alpha}^\textrm{qm}\right)\left(n_\textrm{z} + \frac{1}{2}\right)u^{1/2}\\
    &-\left(\frac{\partial\tilde{\alpha}^\textrm{E1}}{\partial\nu}\delta_\textrm{L} + \frac{3}{2}\tilde{\beta}\left(n_\textrm{z}^2 + n_\textrm{z} + \frac{1}{2}\right)\right)u\\
    &+2\tilde{\beta}\left(n_\textrm{z}+ \frac{1}{2}\right)u^{3/2} - \tilde{\beta}u^2,
\end{split} \label{eq:dnu_s}
\end{equation}
where $h$ is the Plank constant, $u$ is the lattice depth in the units of the recoil energy $E_\textrm{r}$, $\delta_\textrm{L}$ is the detuning of the optical lattice frequency from the exact E1 magic frequency, ${\partial\tilde{\alpha}^\textrm{E1}}/{\partial\nu}$ is the slope of the reduced differential polarizability of the clock transition at the magic wavelength (see the Supplemental Material), $\tilde{\alpha}^\textrm{qm}$ and $\tilde{\beta}$ are the reduced differential E2-M1 polarizability and hyperpolarizability, respectively. 
Notations are the same as in Ref.\,\cite{ushijima2018operational}.

As one can see from Eq.\,(\ref{eq:dnu_s}), nonzero $\tilde{\alpha}^\textrm{qm}$ and $\tilde{\beta}$ result in complex dependence of the frequency shift $\Delta\nu_\textrm{LS}$ on $u$ and $n_\textrm{z}$. 
If $\tilde{\alpha}^\textrm{qm}$ and $\tilde{\beta}$ have the same sign, one can find ``operational'' lattice depth $u^\textrm{op}$ and detuning $\delta^\textrm{op}_\textrm{L}$ for a particular average motional quantum number $n_\textrm{z}^\textrm{avr}$ when this shift is small (ideally close to zero) and is insensitive to small variations of $u^\textrm{op}$ (extremum).
This approach is now implemented in Sr optical lattice clocks \cite{ushijima2018operational}.
However, these operational parameters strongly depend on the average motional quantum number $n_\textrm{z}^\textrm{avr}$ which often requires robust ground-state sideband cooling with $n_\textrm{z}^\textrm{avr} \ll 1$. 
For low trap depths (typically at $u \lesssim 100\,E_\textrm{r}$), sensitivity to $n_\textrm{z}^\textrm{avr}$ is determined by $\tilde{\alpha}^\textrm{qm}$ since the associated clock transition frequency shift scales as $u^{1/2}$.

\begin{figure}[t]
\includegraphics[width=0.5\textwidth]{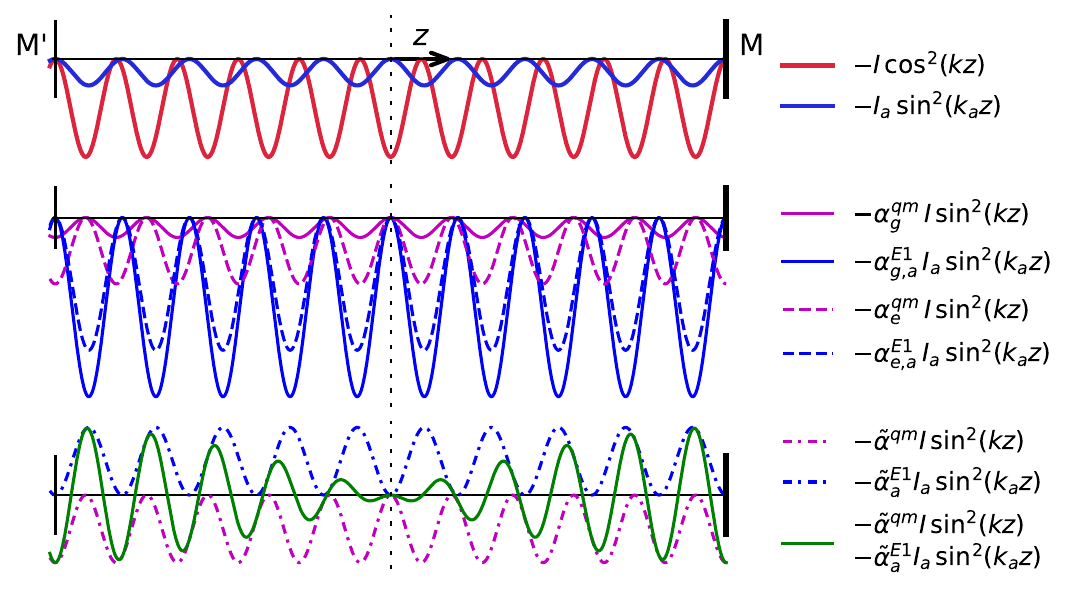}
\caption{\label{fig:lattice} 
Illustration of the idea.
In the central region the auxiliary lattice is spatially shifted from the main lattice by half the period ($\lambda_\textrm{L}/4$).
This leads to similar spatial dependence of its E1-polarizability potential to E2-M1-polarizability potential of the main lattice. 
One can find the detuning and power of the auxiliary lattice (see Eq.\,(\ref{eq:eta_dnu0})) such that its differential E1 potential almost perfectly compensates the differential E2-M1 potential from the main lattice in the central region where the atoms are trapped.
Top - intensity of the main (red) and auxiliary (blue) optical lattices.
Middle - potentials produced by the E2-M1 polarizability in the main lattice (magenta) and by the E1 polarizability in the auxiliary lattice (blue) for the ground (solid) and clock (dashed) levels.
Bottom - difference of these potentials between the clock and ground levels (dash-dotted lines).
The green curve depicts total $\sim\sin^2(kz)$ potential. 
M is the back-reflecting mirror which forms the optical lattices, 
M$'$ is an optional input coupler of the enhancement cavity,
$\tilde{\alpha}^\textrm{E1}_\textrm{a} = \left({\partial\tilde{\alpha}^{E1}}/{\partial\nu}\right)\Delta\nu_\textrm{a}$, other notations are the same as in the text.}
\end{figure}

In this Letter, we propose an approach to eliminate the influence of the E2-M1 differential polarizability $\tilde{\alpha}^\textrm{qm}$ on the clock transition frequency and strongly suppress its sensitivity to the motional state. 
The idea is to use an auxiliary optical lattice, which frequency is detuned by $\Delta\nu_\textrm{a} \equiv \nu_\textrm{a} - \nu_\textrm{L} \sim 1$\,GHz from the main optical lattice frequency, provide its spatial shift of $\lambda_\textrm{L}/4$ relative to the main lattice at the atomic cloud position, and compensate $\tilde{\alpha}^\textrm{qm}$ shift from  the main lattice by nonzero differential E1-polarizability of the auxiliary lattice (see Fig.\,\ref{fig:lattice}). 

In a typical optical lattice clock setup, the optical lattice is formed using either a single back-reflecting mirror or a Fabry-Perot enhancement cavity. 
In the former case, if the auxiliary lattice frequency detuning from the main lattice is $\Delta\nu_\textrm{a} = c/(4 L_\textrm{M})$, where $c$ is the speed of light and $L_\textrm{M}$ is the distance between the atomic cloud and the back-reflecting mirror, the required shift of $\lambda_\textrm{L}/4$ between the nodes of the main and auxiliary lattices occurs in the middle of the atomic cloud.
This is illustrated with red (main lattice) and blue (auxiliary lattice) curves in the upper panel of Fig.\,\ref{fig:lattice}.
The situation is identical for the enhancement cavity, when the atomic cloud is positioned in the middle between the mirrors and $L_\textrm{cav} = 2L_\textrm{M}$.
For a typical $L_\textrm{M} = 50\div150$\,mm, the $\Delta\nu_\textrm{a}=1.5\div0.5$\,GHz.

The combined potential of the main and auxiliary optical lattices is
\begin{equation}
\begin{split}
    U_\xi = -\alpha_\xi^{E1} I \cos^2(kz) - \alpha_\xi^\textrm{qm} I \sin^2(kz) - \beta_\xi I^2 \cos^4(kz) \\
    -\alpha^{E1}_{\xi,\textrm{a}} I_\textrm{a} \sin^2(k_\textrm{a}z) - \alpha^\textrm{qm}_{\xi,\textrm{a}} I_\textrm{a} \cos^2(k_\textrm{a}z) - \beta_{\xi,\textrm{a}} I_\textrm{a}^2 \sin^4(k_\textrm{a}z),
\end{split} \label{eq:u_d0}
\end{equation}
where index ``a'' denotes parameters of the auxiliary optical lattice, and $\lambda_\textrm{L}/4$ spatial shift between the main and auxiliary optical lattices is taken into account at $z=0$. 
Eq.\,(\ref{eq:u_d0}) is exact for orthogonal polarizations of the main and auxiliary lattices and has an extra term if polarizations are parallel, which oscillates at $\Delta\nu_\textrm{a}$. 
If $\Delta\nu_\textrm{a}$ does not match some atomic transition (i.e. hyperfine splitting of the ground or clock levels), influence of this term on the atoms in the ground and clock states and on the clock transition frequency is negligible (see the Supplemental Material).

Since the frequencies of the main and auxiliary lattices are close, we assume below that $\beta_{\xi,\textrm{a}} = \beta_\xi$ and $\alpha^\textrm{qm}_{\xi,\textrm{a}} = \alpha^\textrm{qm}_\xi$, $\xi=g(e)$. 
In the first approximation (see the Supplemental Material), we can put $k_\textrm{a}=k$ and rewrite the potential in Eq.\,(\ref{eq:u_d0}) in the following form:

\begin{equation}
\begin{split}
U_\xi = &-\left(\alpha^{E1}_\xi + \eta_\textrm{a} \alpha^\textrm{qm}_\xi\right) I \cos^2(kz) \\
&- (\alpha^\textrm{qm}_\xi + \eta_\textrm{a} \alpha^{E1}_{\xi,\textrm{a}}) I \sin^2(kz) \\
&- \beta_\xi I^2\left(\cos^4(kz) + \eta_\textrm{a}^2\sin^4(kz)\right),
\end{split} \label{eq:u_d0s}
\end{equation}
where $\eta_\textrm{a} = I_\textrm{a}/I$.
Comparing Eq.\,(\ref{eq:u_d0s}) and Eq.\,(\ref{eq:u_s}), the frequency shift $\Delta\nu_\textrm{LS}^\textrm{mod}$ for such optical potential can be obtained from Eq.\,(\ref{eq:dnu_s}) with the following replacements:
\begin{equation}
\begin{split}
    \frac{\partial\tilde{\alpha}^{E1}}{\partial\nu}\delta_\textrm{L} &\rightarrow \frac{\partial\tilde{\alpha}^{E1}}{\partial\nu}\delta_\textrm{L} + \eta_\textrm{a} \tilde{\alpha}^\textrm{qm} \\
    \tilde{\alpha}^\textrm{qm} &\rightarrow \tilde{\alpha}^\textrm{qm} + \eta_\textrm{a}\frac{\partial\tilde{\alpha}^{E1}}{\partial\nu}\Delta\nu_\textrm{a} \\
    \tilde{\beta}_\textrm{lin} &\rightarrow \tilde{\beta}(1+3/5 \eta_\textrm{a}^2),
\end{split} \label{eq:repls}
\end{equation}
where $\tilde{\beta}_\textrm{lin}$ denotes $\tilde{\beta}$ coefficient in the $u-$linear term in Eq.\,(\ref{eq:dnu_s}) (see the Supplemental Material for a detailed derivation).
As one can see from the second line in (\ref{eq:repls}), the effective $\tilde{\alpha}^\textrm{qm}$ value, and hence the coefficient in $u^{1/2}-$term, can be tuned to a desired value by choosing the detuning $\Delta\nu_\textrm{a}$ and power $\eta_\textrm{a}$ of the auxiliary optical lattice. 
In order to zero the effective $\tilde{\alpha}^\textrm{qm}$ coefficient, the auxiliary lattice parameters are defined by

\begin{equation} \label{eq:eta_dnu0}
    \eta^0_\textrm{a}\Delta\nu_\textrm{a} = -\tilde{\alpha}^\textrm{qm} / ({\partial\tilde{\alpha}^{E1}}/{\partial\nu})
\end{equation}
We note that this condition can always be fulfilled since one can choose any sign of the detuning  $\Delta\nu_\textrm{a}$ of the auxiliary lattice.

As an example, we demonstrate advantages provided by this approach for Sr and Hg optical lattice clocks.
The relevant parameters are listed in Table.\,\ref{tab:pols}.
We assume that the back-reflecting mirror which forms the optical lattice is located at a distance of $L_\textrm{M}=75$\,mm from the atomic cloud position, setting $\Delta\nu_\textrm{a} = \pm1$\,GHz.
It is worth noting that the required $\lambda_\textrm{L}/4$ spatial shift of the auxiliary lattice relative to the main one can be also achieved for $\Delta\nu_\textrm{a} = (1 + 2m)$\,GHz, where $m$ is any integer. 

\begin{table*}
\caption{\label{tab:pols} Relevant parameters of the magic wavelengths for Sr and Hg. 
${\partial\tilde{\alpha}^{E1}}/{\partial\nu}$ is the slope of the differential E1 polarizability, $\delta_\textrm{L}^\textrm{op}$ is the operational detuning of the optical lattice from the exact E1-magic frequency, $\tilde{\alpha}^\textrm{qm}$ and $\tilde\beta$ are E2-M1 polarizability and hyperpolarizability. 
$\eta_\textrm{a}^0$ is the fraction of power in the auxiliary optical lattice to fully compensate $\tilde{\alpha}^\textrm{qm}$ at $|\Delta\nu_\textrm{a}| = 1$\,GHz.
}
\begin{ruledtabular}
\begin{tabular}{cccccccc}
  & $(1/h){\partial\tilde{\alpha}^{E1}}/{\partial\nu}$ & $\delta_\textrm{L}^\textrm{op}$, MHz & 
  $u^\textrm{op}$, $E_\textrm{rec}$ & 
 (1/h)$\tilde{\alpha}^\textrm{qm}$, mHz & 
 (1/h)$\tilde\beta$, \textmu Hz &
 $\eta_\textrm{a}^0$ \\ \hline
 Sr \cite{ushijima2018operational} & $1.735(13)\times10^{-11}$ & $5.3$ & 72 & $-0.962(14)$ & $-0.461(40)$ & 0.055 \\
 Sr \cite{kim2023evaluation} & $1.859(5)\times10^{-11}$ & 0 & 12 &  $-1.24(5)$ & $-0.51(4)$ & 0.067 \\
 Hg \cite{yamanaka2015frequency, katori2015strategies} & $2.1(5)\times10^{-10}$ & -2 & 42 & $11.4$ & $-1.3$ & 0.055 \\
\end{tabular}
\end{ruledtabular}
\end{table*}

\begin{figure}[t]
\includegraphics[width=0.5\textwidth]{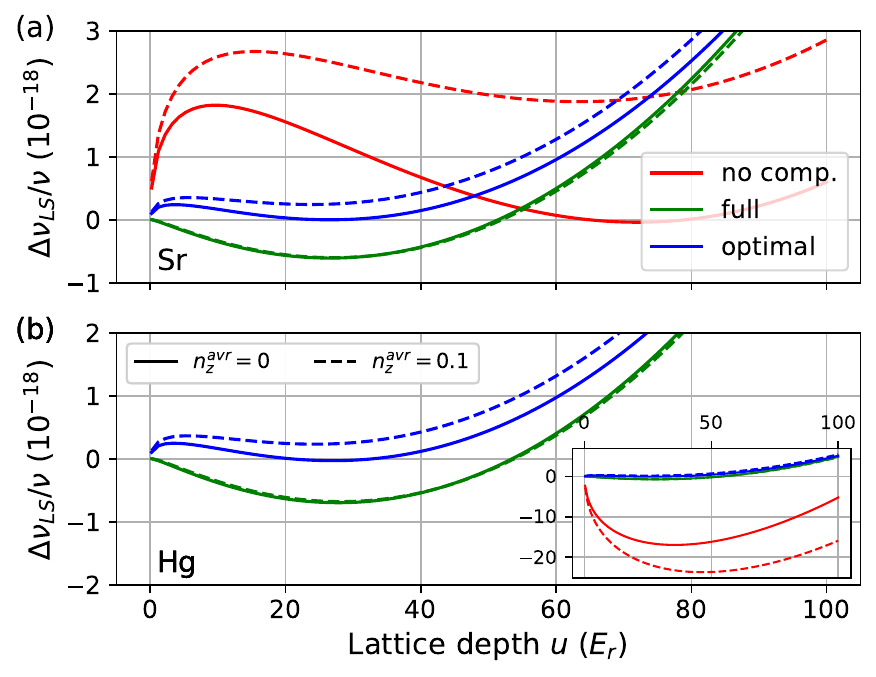}
\caption{\label{fig:SrHg} The lattice-induced relative clock transition frequency shift for Sr (a) and Hg (b).
Solid (dashed) lines correspond to $n_z^\textrm{avr}=0$ ($n_z^\textrm{avr}=0.1$).
Classical configurations (without the auxiliary optical lattice) are shown with red lines.
The green lines represent the light shifts for full compensation of the E2-M1 differential polarizability at $\delta_\textrm{L} = 4.35$\,MHz for Sr and $\delta_\textrm{L} = -2.7$\,MHz for Hg.
The blue lines show the optimal configuration (see text): for Sr the E2-M1 differential polarizability is undercompensated: $\eta_\textrm{a} = 0.8\eta_\textrm{a}^0$ and $\delta_\textrm{L} = 4.3$\,MHz; for Hg it is overcompensated: $\eta_\textrm{a} = 1.045\eta_\textrm{a}^0$ and $\delta_\textrm{L} = -2.72$\,MHz. 
The inset in (b) shows the relative frequency shift for Hg in the larger range (units are the same as in the main plot) to illustrate the much larger uncompensated lattice-induced shift.}
\end{figure}

Figure\,\ref{fig:SrHg} shows the clock transition frequency shift as a function of the main lattice depth $u$ for different auxiliary optical lattice parameters. 
Without the auxiliary lattice (the classical scheme), the clock transition frequency shift for the operational lattice detuning is shown with red curves.
Solid and dashed lines correspond to $n_z^\textrm{avr}=0$ and $n_z^\textrm{avr}=0.1$, respectively.
In order to keep the uncertainty of the clock transition frequency shift below $10^{-18}$ level, one needs to prepare atoms with $n_z^\textrm{avr}\lesssim0.05$ for Sr  and $n_z^\textrm{avr} \lesssim 0.01$ for Hg optical lattice clocks.

In configuration where the auxiliary lattice power is $I_\textrm{a} = \eta^0_\textrm{a} I$ (see Eq.\,(\ref{eq:eta_dnu0}) and Table\,\ref{tab:pols}), one can achieve almost perfect cancellation  of the $n_z^\textrm{avr}$ dependence for $u<100$ (green lines in Fig.\,\ref{fig:SrHg}).
The residual difference between $n_z^\textrm{avr}=0$ and $n_z^\textrm{avr}=0.1$ is below $10^{-19}$ for $u<60$ and is mostly determined by other $n_z^\textrm{avr}$-dependent terms. 
At the operational point, where the lattice-induced shift is insensitive to small variation of $u$, $\Delta\nu_\textrm{LS} \lesssim 10^{-18}$.

Full cancellation of the $\tilde{\alpha}^\textrm{qm}$ coefficient eliminates possibility to find the operational lattice depth where simultaneously $\Delta\nu_\textrm{LS} = 0$ and $\partial(\Delta\nu_\textrm{LS})/\partial u = 0$.
However, this is possible for partial compensation of the $\tilde{\alpha}^\textrm{qm}$.
As an example, for Sr at $\eta_\textrm{a} = 0.8\,\eta_\textrm{a}^0$ (undercompensation) and $\delta_\textrm{L}=4.3$\,MHz the conditions of $\Delta\nu_\textrm{LS} = 0$ and $\partial(\Delta\nu_\textrm{LS})/\partial u = 0$ are fulfilled for $u^\textrm{op'}=25\,E_\textrm{r}$.
In the case of Hg lattice clock, for similar $u^\textrm{op'}$ this can be accomplished at $\eta_\textrm{a} = 1.045\,\eta_\textrm{a}^0$ (overcompensation) and $\delta_\textrm{L}=-2.72$\,MHz.
The corresponding lattice-induced clock transition frequency shifts are shown with blue curves.
In this configuration, in order to keep the clock transition frequency uncertainty below $1\times10^{-19}$ one needs to control $\eta_\textrm{a}$ with 5\% (1\%) accuracy and provide $n_z^\textrm{avr} < 0.05$ ($n_z^\textrm{avr} < 0.01$) for Sr (Hg).
The requirements for Hg are more strict due to 10 times larger E2-M1 differential polarizability $\tilde{\alpha}^\textrm{qm}$.
We note here that the operational lattice depth and detuning, as well as sensitivity to variation of $\eta_\textrm{a}$ and $n_z^\textrm{avr}$, depend on  $\eta_\textrm{a}$, hence one can choose the optimal configuration for each specific case.

For the atoms with a nonzero tensor polarizability, one can get advantage of a possibly large differential tensor polarizability $\Delta\tilde\alpha^{E1}_\textrm{t}$ contribution to the E1-polarizability difference $\Delta\tilde{\alpha}^{E1}_\textrm{a} = ({\partial\tilde{\alpha}^{E1}}/{\partial\nu})\Delta\nu_\textrm{a} + \Delta\tilde{\alpha}^{E1}_\textrm{t}$ by forming the auxiliary lattice with the orthogonal polarization to the main lattice. 
It allows to work with lower power of the auxiliary lattice, however requires that $\tilde\alpha^\textrm{qm}$ and $\Delta\tilde\alpha^{E1}_\textrm{t}$ have different signs.
This may be the best option when the slope of the differential E1-polarizability is small, and either large detuning or high power is required. 
The latter is the case for Tm optical clock \cite{golovizin2021simultaneous} with the optical lattice magic wavelength near $1063.5$\,nm.

We now briefly discuss effects from the spatial shift of the auxiliary lattice nodes relative to the main lattice antinodes due to the difference of the lattices' wavelengths. 
Let's assume that for some main lattice antinode $\kappa=0$ the alignment is ideal.
The spatial shift of the auxiliary lattice nodes relative to the main lattice antinodes is $\delta z^\kappa_\textrm{a} = - \kappa \frac{\Delta\nu_\textrm{a}}{\nu_\textrm{L}}   \frac{\lambda_\textrm{L}}{2}$ and the shift of the potential minimum from the main lattice antinodes is $\delta z^\kappa_\textrm{min} = - \frac{\eta_\textrm{a}}{1- \eta_\textrm{a}} \delta z^\kappa_\textrm{a}$. 
For $|\Delta\nu_\textrm{a}|=1$\,GHz the displacement of the two lattices $\delta z^\kappa_\textrm{a} \sim \kappa\times10^{-12}$\,m.
From the Tailor expansion of Eq.\,(\ref{eq:u_d0}) (a detailed analysis is presented in the Supplemental Material) one finds that:

\begin{itemize}
    \item Linear with $\delta z^\kappa_\textrm{a}$ are only terms proportional to $z$ (and other odd powers).
    The first one is $\Delta U_\textrm{lin} = 2{\eta_\textrm{a}}/(1-\eta_\textrm{a}) ( \frac{\partial\tilde{\alpha}^{E1}}{\partial\nu} \Delta\nu_\textrm{a} I  -2\beta I^2)k^2\,\delta z^\kappa_\textrm{a}\,z$.
    This term is similar to that of the gravity force effect \cite{lemonde2005optical}, however for the parameters of the auxiliary lattice described above, it is $\sim 10^{-6}$ times weaker.
    \item The corrections to the lattice depth and harmonic potential are smaller than the main terms at least by the factor $(k\,\delta z^\kappa_\textrm{a})^2 \approx \kappa^2(\Delta\nu_\textrm{a} / \nu_\textrm{L})^2 \approx 10^{-10} \kappa^2$ and negligible for $|\kappa| \lesssim 1000$ (atomic cloud size of 1\,mm).
    
\end{itemize}

In conclusion, we have shown that the influence from the differential E2-M1 polarizability $\tilde{\alpha}^\textrm{qm}$ on the clock transition frequency can be tuned using the auxiliary low-power optical lattice.
One can either completely cancel this influence or adjust it to the optimal value, significantly suppressing the lattice-induced shift sensitivity to the atomic motional state.
As an example, we show that implementation of this method to Sr and Hg  optical lattice clocks reduces the lattice-induced shift and uncertainty to the low $10^{-19}$ level with very realistic requirements on the parameters control.
For Hg optical clock, this is more than two orders of magnitude improvement. 

The ability to tune the $\tilde{\alpha}^\textrm{qm}$ coefficient allows one to achieve $\Delta\nu_\textrm{LS}=0$ and ${\partial(\Delta\nu_\textrm{LS}})/{\partial u}=0$ at any required lattice depth $u$.
One can significantly reduce the clock transition frequency shift and uncertainty at the ``magic'' depth $u_\textrm{M}\approx10\,E_\textrm{r}$ in Sr optical clock at which the density shift is suppressed to below $10^{-19}$ \cite{aeppli2022hamiltonian} (see the Supplemental Material).
Owing to its generality, this method is applicable to any other atomic element and especially useful when the E2-M1 differential polarizability is large (as in Hg optical clock).
Experimental implementation is quite straightforward -- one can produce a single frequency sideband with an electro-optic modulator \cite{loayssa2004optical} in the existing optical lattice beam with no other modifications since the main and auxiliary lattice beams' paths would be the same.

We thank D.\,Tregubov, I.\,Zalivako and N.\,Kolachevsky for valuable discussions and careful readings of the manuscript.
This work was supported by RSF grant no. 21-72-10108.

\bibliography{lib}

\end{document}



\title{Supplemental Material to \\ Compensation of the maltipolar polarizability shift in optical lattice clocks}

\author{Artem Golovizin}
\email{artem.golovizin@gmail.com}
\affiliation{P.N.\,Lebedev Physical Institute, Leninsky prospekt 53, 119991 Moscow, Russia}
\affiliation{Russian Quantum Center, Bolshoy Bulvar 30,\,bld.\,1, Skolkovo IC, 121205 Moscow, Russia}

\date{\today}

\maketitle


\subsection{\label{subsec:full}Potential produced by two optical lattices}

Let's consider two plane waves, main and auxiliary, with the wavevectors $\mathbf{k}=k\,\mathbf{\hat{z}}$ (here $\mathbf{\hat{z}}$ is a unit vector) and $\mathbf{k}_\textrm{a} = k_\textrm{a}\,\mathbf{\hat{z}}$ with electric field $\mathbf{E}\cos(\omega t - k z)$ and $\mathbf{E}_\textrm{a}\cos(\omega_\textrm{a} t - k_a z)$, which are reflected back by the mirror at $z=0$.
The produced intensity is

\begin{equation}\label{eq:summing}
    \begin{split}
        I &= c\epsilon_0\left<\left(\mathbf{E} \cos(\omega t + k z) + \mathbf{E}_a \cos(\omega_a t + k_a z) + \mathbf{E} \cos(\omega t - k z) + \mathbf{E}_a \cos(\omega_a t - k_a z)\right)^2\right> \\
        &= c\epsilon_0\left<\left(2 \mathbf{E} \cos(k z) \cos(\omega t) + 2 \mathbf{E}_a \cos(k_a z) \cos(\omega_a t)\right)^2\right> \\
        &= c\epsilon_0\left<4 E^2 \cos^2(k z) \cos^2(\omega t) + 4 E_a^2 \cos^2(k_a z) \cos^2(\omega_a t) + 4 \mathbf{E}\, \mathbf{E}_a \cos(k z) \cos(k_a z) \left(\cos((\omega_a - \omega) t) + \cos((\omega_a + \omega) t) \right)\right> \\
        &= I \cos^2(k z) + \eta_a I \cos^2(k_a z) + 2 \sqrt{\eta_a} I (\mathbf{e}\mathbf{e}_a) \cos(k z) \cos(k_a z)\cos(\Delta\omega_a t)
    \end{split}
\end{equation}
where we performed time averaging of oscillations at optical frequencies, $I = 4\times c\epsilon_0 E^2/2$ and $I_a = 4\times c\epsilon_0 E_a^2/2$, $\eta_a = I_a/I$, $\Delta\omega_a \equiv 2\pi \Delta\nu_a = \omega_a - \omega$ and $\mathbf{e}$ and $\mathbf{e}_a$ are the polarization vectors of the main and auxiliary latices, respectively.
In the case of perpendicular polarizations, $(\mathbf{e}\mathbf{e}_a) = 0$ and the total intensity is the sum of two independent standing waves.
If the polarizations are parallel, $(\mathbf{e}\mathbf{e}_a) = 1$ and the last term is nonzero.
The atoms are to be located where the nodes of the additional lattice coincide with the antinodes of the main lattice $k_a z \approx k z + \pi/2$, which occurs at $z_w = \frac{\pi}{2(k_a - k)} = \frac{c}{4\Delta\nu_\textrm{a}}$ (equals to $L_\textrm{m}$ - the distance between the atoms and the back-reflecting mirror introduced in the main text).
One can rewrite the last term in Eq.\,\ref{eq:summing} for $z'$ near $z_w$ in the following form:

\begin{equation}
    \begin{split}
        I_3 \approx 2 \sqrt{\eta_a} I \cos(k z') \sin(k z') \cos(\Delta \omega_a t)
    \end{split}
\end{equation}
Since the atoms are localized near the main lattice antinodes, one can consider $kz' = \kappa \pi + z''$ with $z'' \ll \lambda/2$, then  $\cos(k z') \sin(k z') \approx k z''$.
Average of $k z''$ is near 0 and the upper bound is $k z_0 = \eta$, where $z_0$ is the spread of the zero-point wavefunction and $\eta$ is the Lamb-Dicke parameter.
Thus amplitude of the oscillating potential is less than $U_{osc}^{max} = 2\sqrt{\eta_a} \eta U_0 \sim 10^{-2}\dots10^{-1} U_0$, where $U_0$ is the main lattice depth.
$\Delta\nu_a \sim 1$\,GHz is much larger than the motional frequencies of the atoms in the trap, so no heating of atoms is expected.
For large enough detuning of $\Delta\nu_a$ from the hyperfine-structure splittings of the ground and clock levels, the Raman transitions shouldn't be excited as well.
Hence, this term averages to zero on a timescale of more than 1\,\textmu s  and has negligible influence on the atoms and the clock transition frequency.

\subsection{Effect of displacement of the two optical lattices}\label{sec:pot_shift}

\begin{figure}[h]
\includegraphics[width=0.7\textwidth]{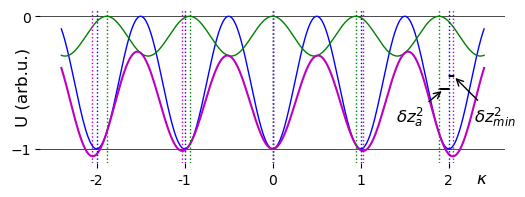}
\caption{\label{fig:lattice} Illustration of the potential U formed by two optical lattices near the node number $\kappa=0$.
Blue solid line shows the main lattice, green line shows the auxiliary lattice with $\eta_\textrm{a}=0.3$ and $\nu_\textrm{a} = 1.06\,\nu_\textrm{L}$.
Magenta solid line shows the total potential. 
Dotted lines indicate positions of the antinodes of the main lattice, nodes of the auxiliary lattice and coordinates of the total potential minimums.}
\end{figure}

The full potential produced by two standing waves (optical lattices) is described by Eq.\,3 from the main text and is shown with magenta line in Fig.\,\ref{fig:lattice} for $\Delta\nu_\textrm{a} / \nu_\textrm{L} = 0.06$ and $\eta_\textrm{a} = 0.3$ (for illustration purposes). 
In this section we omit notation of the level ($\xi$, $g$ or $e$ in the main text) for simplicity.
We assume that for a some antinode of the main lattice at a distance of $z_w \equiv L_\textrm{m}$ from the back-reflecting mirror the auxiliary lattice  has exactly a node: we designate this antinode with $\kappa=0$ and set $z=0$ (note that $z, z'$ and $z''$ here are different than in Sec.\,\ref{subsec:full}). 
The spatial shift of the auxiliary lattice node relative to the main lattice antinode is $\delta z^\kappa_\textrm{a} = - \kappa \lambda_\textrm{L}/2 \times  \Delta\nu_\textrm{a} / \nu_\textrm{L}$, which is of the order of $\kappa \times 10^{-12}$\,m for $|\Delta\nu_\textrm{a}| = 1$\,GHz.
Since $\eta_\textrm{a} < 1$ (the auxiliary lattice intensity $I_\textrm{a}$ is smaller than the main lattice intensity $I$), the potential minimum in this antinode is at the same coordinate as for the main lattice (see Fig.\,\ref{fig:lattice}). 
In order to find the displacement of the potential minimum for other antinodes relative to the main lattice, we keep only E1-polarizability terms in Eq.\,3 from the main text, additionally assume $\alpha^{E1}_\textrm{a} \approx \alpha^{E1}$ and substitute $z = n\lambda_\textrm{L}/2 + z'$ and $k_\textrm{a} = k + \Delta k$, where $z'\approx 0$ for each antinode and $\Delta k = k\Delta\nu_\textrm{a}/\nu_\textrm{L}$. 
The trap potential for the $\kappa$'s antinode is

\begin{equation}
\begin{split}
    U^\kappa = -\left[\cos^2(k(\kappa\lambda_\textrm{L}/2 + z'))
    +\eta_\textrm{a}\sin^2((k + \Delta k)(\kappa\lambda_\textrm{L}/2 + z'))\right] \alpha^{E1} I  \\
    \approx -\left[\cos^2(k\,z') + \eta_\textrm{a}\sin^2(k (z' - \delta z^\kappa_\textrm{a}))\right] \alpha^{E1} I
\end{split} \label{eq:u_pot}
\end{equation}

where we used that $\cos^2(k (\kappa\lambda_\textrm{L}/2 + z')) = \cos^2(\kappa \pi+k\,z') = \cos^2(k\,z')$.
Position of the potential minimum $\delta z^\kappa_\textrm{min}$ relative to the main lattice antinode can be found by performing the Tailor expansion of Eq.\,\ref{eq:u_pot}:
\begin{equation} \label{eq:u_tailor_min}
    U^{\kappa,0} = -[1 - (k\,z')^2 + \eta_\textrm{a}\left(k (z' - \delta z^\kappa_\textrm{a})\right)^2] \alpha^{E1} I
\end{equation}
which minimum is at
\begin{equation}
    \delta z^\kappa_\textrm{min} = -\frac{\eta_\textrm{a}}{1- \eta_\textrm{a}} \delta z^\kappa_\textrm{a}
\end{equation}

Now we can rewrite Eq.\,3 from the main text for the $\kappa$'s antinode substituting $z = \kappa\lambda_\textrm{L}/2 + \delta z^\kappa_\textrm{min} + z''$, where $z''=0$ at the potential minimum for each antinode:

\begin{equation}
\begin{split}
    U^\kappa = -\alpha^{E1} I \cos^2(k(\delta z^\kappa_\textrm{min} + z'')) - \alpha^\textrm{qm} I \sin^2(k(\delta z^\kappa_\textrm{min} + z'')) - \beta I^2 \cos^4(k(\delta z^\kappa_\textrm{min} + z'')) \\
    -\alpha^{E1}_\textrm{a} I_\textrm{a} \sin^2(k(\delta z^\kappa_\textrm{min} -  \delta z^\kappa_\textrm{a} + z'')) - \alpha^\textrm{qm} I_\textrm{a} \cos^2(k(\delta z^\kappa_\textrm{min} -  \delta z^\kappa_\textrm{a} + z'')) \\ 
    - \beta I_\textrm{a}^2 \sin^4(k(\delta z^\kappa_\textrm{m} -  \delta z^\kappa_\textrm{a} + z''))
\end{split} \label{eq:u_d_full}
\end{equation}

Implementing the Tailor expansion on $z''$ and $\delta z^n_\textrm{a}$ we can rewrite Eq.\,\ref{eq:u_d_full} in the form
\begin{equation}
    U^\kappa = \sum_{i=0} c^\kappa_i \, (k z'')^i
\label{eq:u_d_sum}
\end{equation}
where the first 3 coefficients are

\begin{subequations}
\begin{align}
    c^\kappa_0 &= - (\alpha^{E1} + \eta_\textrm{a} \alpha^\textrm{qm})I - \beta I^2 \nonumber \\ 
    &- \frac{\eta_\textrm{a}}{(1-\eta_\textrm{a})^2} \left((\alpha^{E1}_\textrm{a} - \eta_\textrm{a} \alpha^{E1} - (1-\eta_\textrm{a})\alpha^\textrm{qm}) I - 2\eta_\textrm{a}\beta I^2\right)(k\,\delta z^\kappa_\textrm{a})^2 + O\left((\delta z^\kappa_\textrm{a}/\lambda_\textrm{L})^4\right)
    \label{eq:u_coeff0} \\
    c^n_1 &= 2\frac{\eta_\textrm{a}}{1-\eta_\textrm{a}} ( \Delta\alpha^{E1}_\textrm{a} I  -2\beta I^2)   (k\,\delta z^n_\textrm{a}) + O((\delta z^n_\textrm{a}/\lambda_\textrm{L})^3) 
    \label{eq:u_coeff1} \\ 
    c^n_2 &= \left(\alpha^{E1}- \alpha^\textrm{qm} - \eta_\textrm{a} (\alpha^{E1}_\textrm{a} - \alpha^\textrm{qm})\right)I  + 2\beta I^2 \nonumber \\ 
    &+ 2\frac{\eta_\textrm{a}}{(1-\eta_\textrm{a})^2}\left((\alpha^{E1}_\textrm{a} - \alpha^\textrm{qm} - \eta_\textrm{a} (\alpha^{E1} - \alpha^\textrm{qm}) I - 8\eta_\textrm{a}\beta I^2 \right)(k\,\delta z^n_\textrm{a})^2 + O(
    (\delta z^n_\textrm{a}/\lambda_\textrm{L})^4) \label{eq:u_coeff2}
\end{align} 
\end{subequations}

As we pointed out in the main text, the relative value of the $\kappa$-dependent terms in coefficients $c^\kappa_0$ (optical lattice depth, Eq.\,\ref{eq:u_coeff0}) and  $c^\kappa_2$ (harmonic potential, Eq.\,\ref{eq:u_coeff2}) are of the order of $\eta_\textrm{a}(k\,\delta z^\kappa_\textrm{a})^2 \approx 10^{-10} \kappa^2$
This corrections are of the same order as due to the intensity change from the divergence of the gaussian beam  with the beam waist $w_0 = 100$\,\textmu m.

\subsection{\label{subsec:decomposition} Modification of the lattice-induce clock transition frequency shift} \label{sec:shift_all}

Here we derive the lattice-induced clock transition frequency shift $\Delta\nu_\textrm{LS}^\textrm{mod}$ in the presence of the auxiliary optical lattice.
Performing the Tailor expansion of Eq.\,4 from the main text near $z=0$ up to $z^4$ one obtains
\begin{equation}
\begin{split}
U_\xi = &-(\alpha_\xi^{E1} +\eta_\textrm{a} \alpha_\xi^\textrm{qm})I - \beta_\xi I^2 \\
& + \left[\left((\alpha_\xi^{E1} + \eta_\textrm{a} \alpha_\xi^\textrm{qm})-(\alpha_\xi^\textrm{qm} + \eta_\textrm{a} \alpha^{E1}_{\xi,\textrm{a}}) \right)I + 2\beta_\xi I^2 \right] (k\,z)^2 \\
&- 1/3 \left[\left((\alpha_\xi^{E1} + \eta_\textrm{a} \alpha_\xi^\textrm{qm})-(\alpha_\xi^\textrm{qm} + \eta_\textrm{a} \alpha^{E1}_{\xi,\textrm{a}}) \right)I + 5(1 + 3/5\eta_\textrm{a}^2)\beta_\xi I^2\right](k\,z)^4
\end{split} \label{eq:u_d_tailor}
\end{equation}
where $\xi$ denotes either ground $g$ or clock $e$ levels.
We note that correction to the hyperpolarizability takes place only in $z^4$-term. 
The lattice-induced clock transition frequency shift for $n_z$-th motional state $\ket{n_z}$ can be found from 
\begin{equation}
     h\Delta\nu_\textrm{LS}(I,n_\textrm{z}) \approx \bra{n_\textrm{z}}{U_e(I) - U_g(I)}\ket{n_\textrm{z}}.
\end{equation}
Introducing reduced differential polarizabilities and the lattice depth $u$ in the units of the recoil energy $E\textrm{r}$ similar to  Ref.\,\cite{ushijima2018operational}
\begin{equation}\label{eq:notations}
    \begin{split}
        \tilde{\alpha}^{E1} &= (\alpha_e^{E1}-\alpha_g^{E1}) \frac{E_\textrm{r}}{\alpha_g^{E1}} \\
        \tilde{\alpha}^\textrm{qm} &= (\alpha_e^\textrm{qm}-\alpha_g^\textrm{qm}) \frac{E_\textrm{r}}{\alpha_g^{E1}} \\
        \tilde{\beta} &= (\beta_e-\beta_g) \left(\frac{E_\textrm{r}}{\alpha_g^{E1}}\right)^2 \\
        u &= I \left(\frac{E_\textrm{r}}{\alpha_g^{E1}}\right)^{-1}
    \end{split}
\end{equation}

the modified clock transition frequency shift is
\begin{equation}
\begin{split}
    h\Delta\nu^\textrm{mod}_\textrm{LS}(u,\delta_\textrm{L},n_\textrm{z}) &= \left((\frac{\partial\tilde{\alpha}^{E1}}{\partial\nu}\delta_\textrm{L} + \eta_\textrm{a} \tilde{\alpha}^\textrm{qm}) - (\tilde{\alpha}^\textrm{qm} + \eta_\textrm{a}\frac{\partial\tilde{\alpha}^{E1}}{\partial\nu}\Delta\nu_\textrm{a})\right)\left(n_\textrm{z} + \frac{1}{2}\right)u^{1/2}\\
    &-\left((\frac{\partial\tilde{\alpha}^{E1}}{\partial\nu}\delta_\textrm{L} + \eta_\textrm{a} \tilde{\alpha}^\textrm{qm}) + \frac{3}{2}(1+3/5 \eta_\textrm{a}^2)\tilde{\beta}\left(n_\textrm{z}^2 + n_\textrm{z} + \frac{1}{2}\right)\right)u\\
    &+2\tilde{\beta}\left(n_\textrm{z}+ \frac{1}{2}\right)u^{3/2} - \tilde{\beta}u^2,
\end{split} \label{eq:dnu_s}
\end{equation}

\subsection{Lattice-induced clock transition frequency shift at the ``magic'' lattice depth}

\begin{figure}[h]
\includegraphics[width=0.7\textwidth]{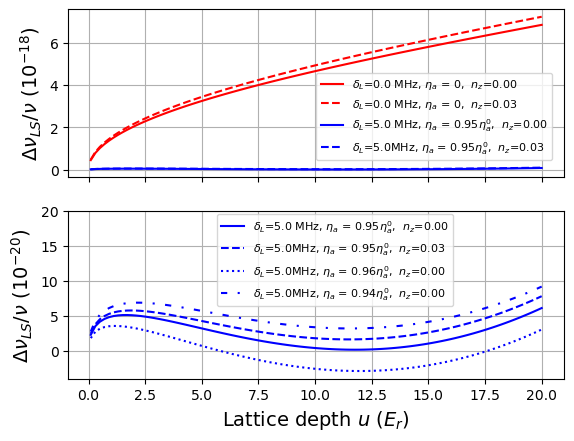}
\caption{\label{fig:low_depth} Lattice-induced clock transition frequency shift as a function of the lattice depth near the ``magic'' depth $u_\textrm{M}=10\,E_\textrm{r}$. 
Red lines depict the shifts without compensation (``classical'' scheme), blue lines - with the auxiliary lattice. 
Solid (dashed) lines correspond to $n_\textrm{z}^\textrm{avr}=0$ ($n_\textrm{z}^\textrm{avr}=0.03$).
Here $\eta_\textrm{a}^0 = 0.67$ is calculated based on data in Ref.\,\cite{kim2023evaluation}.
Dotted and dash-dotted lines in the bottom plot depict influence of 1\% variation of $\eta_\textrm{a}$ on the clock transition frequency shift.}
\end{figure}

In Ref.\,\cite{aeppli2022hamiltonian}, the authors found ``magic'' optical lattice depth $u_\textrm{M}\approx10\,E_\textrm{r}$ for Sr optical clock at which the density shift is strongly suppressed owing to compensation of the off-site $s-$wave and on-site $p-$wave collisional shift. 
In Ref.\,\cite{kim2023evaluation} a detailed analysis of the lattice-induced clock transition frequency shift was performed. 
For $u_\textrm{M}=10\,E_\textrm{r}$ and optical lattice frequency equal to the E1-magic frequency ($\delta_\textrm{L}=0$), its uncertainty was evaluated to be $3.5\times10^{-19}$. 
Solid red line on Fig.\,\ref{fig:low_depth} shows corresponding lattice-induced clock transition frequency shift as a function of the lattice depth for $n_\textrm{z}^\textrm{avr}=0$ (red dashed line for $n_\textrm{z}^\textrm{avr}=0.03$).
When adding the auxiliary optical lattice with $\eta_\textrm{a} = 0.95\eta_\textrm{a}^0$ and $\Delta\nu_\textrm{a} = 1$\,GHz, the lattice light shift can be tuned to zero at $u_\textrm{M}$ with the uncertainty from $n_\textrm{z}^\textrm{avr}=0(0.03)$ at $10^{-20}$.
The detuning of the main lattice from the exact E1 magical frequency is $\delta_\textrm{L}=5.0$\,MHz .

\bibliography{lib}